# Titrating Polyelectrolytes – Variational Calculations and Monte Carlo Simulations

Bo Jönsson[1] and Magnus Ullner[2]

Physical Chemistry 2, Chemical Center, University of Lund
Box 124, S-221 00 Lund, Sweden

Carsten Peterson[3], Ola Sommelius[4] and Bo Söderberg[5]

Department of Theoretical Physics, University of Lund
Sölvegatan 14A, S-22362 Lund, Sweden

Abstract:

Variational methods are used to calculate structural and thermodynamical properties of a titrating polyelectrolyte in a discrete representation. In the variational treatment, the Coulomb potentials are emulated by harmonic repulsive forces between *all* monomers; the force constants are used as variational parameters. The accuracy of the variational approach is tested against Monte Carlo data. Excellent agreement is obtained for the end-to-end separation and the apparent dissociation constant for the unscreened Coulomb chain. The short-range screened Coulomb potential is more difficult to handle variationally and its structural features are less well described, although the thermodynamic properties are predicted with the same accuracy as for the unscreened chain. The number of variational parameters is of the order of $N^2$, where $N$ is of the number of monomers, and the computational effort scales like $N^3$. In addition, a simplified variational procedure with only two parameters is pursued, based on a rigid-rod approximation of the polymer. It gives surprisingly good accuracy for certain physical properties.

[1] fk2boj@grosz.fkem2.lth.se
[2] fk2mul@dix.fkem2.lth.se
[3] carsten@thep.lu.se
[4] ola@thep.lu.se
[5] bs@thep.lu.se

# 1 Introduction

With increasing computer resources and refined algorithms it has now become possible to investigate structural and thermodynamic properties of polymer chains with several thousand monomeric units using simulation techniques. In particular, charged polymers have recently received an increased interest and a number of simulation studies have appeared in the literature [1, 2, 3, 4, 5, 6, 7, 8]. Most polyelectrolytes will under normal solution conditions exhibit an acid-base equilibrium, i.e. titratable groups in a polymer will exchange protons with the solution and the polymer net charge will vary as a function of the solution pH. This extra degree of freedom makes polyelectrolytes particularly versatile in many technical and biological applications [9, 10, 11, 12, 13, 14, 15]. In simulations, this phenomenon can be described by coupling the polyelectrolyte to an external proton bath of fixed chemical potential or pH; such grand canonical Monte Carlo simulations have been performed by several groups [16, 17, 18, 19].

Many polyelectrolytes undergo a dramatic structural change upon a successive ionization, i.e. the Coulombic repulsion between charged monomeric units forces the chain to adopt more extended conformations. This is particularly dramatic at low polymer and salt concentration. Sometimes the Coulombic repulsion is counteracted by attractive interactions leading to a helix-coil transition as a function of pH [20, 21, 22, 23]. In hydrophobically modified polyelectrolytes the attractive interactions tend to dominate and rather modest structural changes are seen upon ionization. Proteins are a special class of polyelectrolytes, which, although they may denature, still show rather small structural changes when going from the isoelectric point to either high or low pH. The competition between electrostatic monomer-monomer repulsion and specific attractive interactions has also been studied with simulation techniques [24, 17].

So far most polyelectrolyte studies have focussed on the behaviour of a single chain at infinite dilution. Additional salt has been introduced in an approximate way via screened Coulomb interactions. Finite polymer concentration and explicit salt particles impose considerable constraints on the chain lengths to be handled. A few studies, however, of non-titrating polyelectrolytes with explicit salt particles have been presented, but then only for fairly short chains [4, 25, 26, 27, 28]. Simulations with finite polymer concentration have also been carried through, but such studies are at present only feasible for a limited chain length [27, 28]. In this paper the accuracy of less computer demanding non-stocahstic approaches are investigated using Monte Carlo simulations.

A number of approximate models for titrating polyelectrolytes have been suggested focussing on different phenomena and applying tools of varying mathematical sophistication. Existing theories, however, are seldom based on models that incorporate both a fully flexible chain and a discrete representation of the monomers. Here we suggest a variational approach, which has given excellent results for non-titrating polyelectrolytes [29, 30], based on a discrete representation of the polymer chain. That is, each monomeric unit is connected to its neighbours with a harmonic bond and may carry a unit charge depending on the solution pH. We apply the variational technique to a linear chain although the method is perfectly general and can be applied to any chain topology. The variational solution to the titrating chain is obtained for different chain lengths and salt concentrations and the results are compared to results obtained via Monte Carlo simulations.

The full variational solution is obtained at a much lower computational effort than what is required by simulations. As a further advantage it also provides the chain free energy. Still, the numerical procedure leading to the variational result is non-trivial, and we have also developed a simpler, but of course less accurate, variational scheme based on only two variational parameters.

This paper is organized as follows. In Sect. 2 we describe the details of the model we use for the titrating polymer. The variational technique of refs. [29, 30] is briefly reviewed in Sect. 3 together with some new numerical results for large $N$ for non-titrating chains. In Sect. 4 the variational formalism is generalized to titrating chains, while the details of the Monte Carlo calculations can be found in Sect. 5. Sect. 6 contains the corresponding comparisons between Monte Carlo and variational results for chains in a salt-free environment as well as in salt solutions. Finally a brief summary and our conclusions can be found in Sect. 7.

## 2  The Model

The polyelectrolyte is regarded as an infinitely diluted polyacid in aqueous solution. The monomers form a linear chain where each monomer represents a titrating site that can be either protonated or deprotonated, i.e. be uncharged or carry one unit of negative charge. The chain is freely jointed, with neighbouring monomers connected by harmonic bonds. The implicit assumption is that the part of the underlying neutral polymer backbone that separates the neighbouring charged groups obeys a Gaussian distribution for its end-to-end separation. This is a reasonable approach for a chain where the separation of neighbouring charges is much larger than the persistence length of the underlying neutral chain.

The solvent is treated as a dielectric continuum with a permittivity equal to that of water at room temperature. It also acts as a proton reservoir via a chemical potential given by

$$\tilde{\mu} = k_B \tilde{T} \ln 10 \, (pK_0 - pH) \qquad (1)$$

where $pH$ is that of the bulk, and $pK_0$ is the intrinsic $pK_a$ of a monomer. In the case of a salt solution, an additional effect of the solvent is a Debye-Hückel screening of the electrostatic interactions between all charged monomers, and the total interaction energy for $N$ monomers becomes

$$\tilde{E} = \tilde{E}_G + \tilde{E}_C + \tilde{E}_\mu = \frac{k}{2} \sum_{i \neq N} |\tilde{\mathbf{x}}_{i,i+1}|^2 + \frac{e^2}{4\pi\epsilon_r\epsilon_0} \sum_i \sum_{j>i} \frac{Z_i Z_j \, exp(-\tilde{\kappa}|\tilde{\mathbf{x}}_{ij}|)}{|\tilde{\mathbf{x}}_{ij}|} - \tilde{\mu} \sum_i Z_i \qquad (2)$$

where $\tilde{\mathbf{x}}_{ij}$ is the distance between monomer $i$ and $j$, $e$ is the electronic charge, $\epsilon_r$ is the dielectric constant of the solution (78.3 in all simulations), $\epsilon_0$ is the permittivity of vacuum and $Z_i$ the amount of charge on monomer $i$ (either 0 or $-1$). We use the tilde notation $\tilde{E}$, $\tilde{\mathbf{x}}_i$, etc. for physical quantities in conventional units, and reserve $E$, $\mathbf{x}_i$, etc. for dimensionless ones, which will be used in the variational formalism below. The force constant, $k$, is implicitly given through the input parameter $\tilde{r}_0 = (e^2/4\pi\epsilon_r\epsilon_0 k)^{1/3}$, which is the equilibrium distance for a fully charged dimer and it is set to 6 Å in all calculations. In the case of a $1:-1$ salt $\tilde{\kappa}$ is given by $\tilde{\kappa} = (2e^2 c_s/\epsilon_r\epsilon_0 N_A k_B \tilde{T})^{1/2}$, where $c_s$ is the salt concentration (in mM), $k_B$ is the Boltzmann constant, $N_A$ is the Avogadro number and $\tilde{T}$ is the temperature.

## 2.1 Dimensionless Formulation with Relative Coordinates

Using dimensionless coordinates $\mathbf{x}_i$, given by $\tilde{\mathbf{x}}_i = r_0 \mathbf{x}_i$, we may define a rescaled temperature $T$,

$$T = \frac{k_B \tilde{T}}{k r_0^2} \tag{3}$$

and a likewise rescaled chemical potential $\mu$,

$$\mu = \frac{\tilde{\mu}}{k r_0^2} \tag{4}$$

The negative exponent of the Boltzmann factor can then be written as

$$\frac{\tilde{E}}{k_B \tilde{T}} = \frac{E}{T} \tag{5}$$

with the rescaled energy given by

$$E = \frac{1}{2} \sum_i |\mathbf{x}_{i,i+1}|^2 + \sum_{i<j} \frac{s_i s_j e^{-\kappa |\mathbf{x}_{ij}|}}{|\mathbf{x}_{ij}|} + \mu \sum_i s_i \tag{6}$$

where we have replaced $Z_i$ by more convenient $\{0,1\}$ variables $s_i = -Z_i$.

In what follows, relative coordinates will mostly be used; instead of the absolute monomer positions $\mathbf{x}_i$, the *bond vectors* $\mathbf{r}_i$,

$$\mathbf{r}_i \equiv \mathbf{x}_{i+1} - \mathbf{x}_i, \quad i = 1, \ldots, N-1 \tag{7}$$

will be considered the fundamental variables. In this way complications due to the translational zero-modes are avoided; in addition, the convergence of the variational algorithm to be described below becomes considerably faster, especially at high temperatures. The energy of the chain then takes the form

$$E(\mathbf{r}, s) = E_G + E_C + E_\mu = \frac{1}{2} \sum_{i=1}^{N-1} \mathbf{r}_i^2 + \sum_\sigma \frac{s_{l_\sigma} s_{r_\sigma} e^{-\kappa r_\sigma}}{r_\sigma} + \mu \sum_i s_i \tag{8}$$

where $\sigma$ runs over contiguous non-nil sub-chains, with

$$\mathbf{r}_\sigma \equiv \sum_{i \in \sigma} \mathbf{r}_i \tag{9}$$

corresponding to the distance vector between the endpoints, $l_\sigma$ and $r_\sigma$, of the subchain.

## 2.2 The Apparent Equilibrium Constant

The average degree of dissociation, $\alpha \equiv \frac{1}{N} \sum_i \langle s_i \rangle$, can be interpreted in terms of an effective chemical equilibrium constant, $pK$

$$pK = pH - \lg\left(\frac{\alpha}{1-\alpha}\right) \tag{10}$$

Without the interaction, we would have

$$\frac{\alpha}{1-\alpha} = \exp\left(-\frac{\mu}{T}\right) \tag{11}$$

A simple measure of the effect of the interaction on the chemical balance is then given by the change in $pK$:

$$\Delta pK = pK - pK_0 = -\frac{\mu}{T \ln 10} - \lg \frac{\alpha}{1-\alpha} \tag{12}$$

# 3 Variational Treatment of Conformational Degrees of Freedom

## 3.1 Generic Formalism

In this section we will focus on the conformational degrees of freedom, ignoring the titration; thus we consider a non-titrating polymer, with an energy obtained from eq. (8) by setting $s_i = 1$ and disregarding the $\mu$ term:

$$E(\mathbf{r}) = E_G + E_C = \frac{1}{2} \sum_{i=1}^{N-1} \mathbf{r}_i^2 + \sum_\sigma \frac{e^{-\kappa r_\sigma}}{r_\sigma} \tag{13}$$

A suitable variational approach to such a system is based on an effective energy Ansatz [31, 32, 29, 30]

$$E_V/T = \frac{1}{2} \sum_{ij} G^{-1}_{ij} (\mathbf{r}_i - \mathbf{a}_i) \cdot (\mathbf{r}_j - \mathbf{a}_j) \tag{14}$$

where $\mathbf{a}_i$ define average bond vectors, around which Gaussian fluctuations are allowed, described by the symmetric, positive-definite correlation matrix $G_{ij}$, the matrix inverse of which appears in the energy.

Using this effective energy, the exact free energy $F = -T \ln Z$ of the polymer is approximated from above [33] by the variational one

$$\hat{F} = -TS_V + \langle E \rangle_V \geq F \tag{15}$$

where $S_V$ is the variational entropy, and $\langle E \rangle_V$ is the average of the true energy in the trial Boltzmann distribution $\propto \exp(-E_V/T)$. The parameters $G_{ij}$ and $\mathbf{a}_i$ are to be determined so as to minimize the variational free energy $\hat{F}$. The resulting effective Boltzmann distribution is then used to approximate expectation values $\langle \cdots \rangle$ by effective (variational) ones $\langle \cdots \rangle_V$. Thus, we have e.g. $\langle \mathbf{r}_i \rangle_V = \mathbf{a}_i$ and $\langle \mathbf{r}_i \cdot \mathbf{r}_j \rangle_V = \mathbf{a}_i \cdot \mathbf{a}_j + 3G_{ij}$. For potentials diverging like $1/r^3$ or worse at short distances, $\langle E \rangle_V$ will be divergent, and the approach breaks down. However, such potentials are not physical.

At high $T$ and/or small $N$, the resulting $\mathbf{a}_i$ will vanish. By setting $\mathbf{a}_i = 0$ in eq. (14), a restricted Ansatz is obtained, that in the screened case yields better results numerically than the $\mathbf{a}_i \neq 0$ case; this restricted version will be frequently used below.

The minimization of $\hat{F}$ with respect to $G_{ij}$ and $\mathbf{a}_i$ gives rise to a set of matrix equations to be solved iteratively. These are considerably simplified, and the symmetry and positivity constraints on $G_{ij}$ are automatic, if $G_{ij}$ is expressed as the product of a matrix and its transpose:

$$G_{ij} = \sum_{\mu=1}^{N-1} z_{i\mu} z_{j\mu} = \mathbf{z}_i \cdot \mathbf{z}_j \tag{16}$$

The interpretation of the local parameter $\mathbf{z}_i$ is simple – it is a fluctuation amplitude for the $i$th bond vector $\mathbf{r}_i$. We can write

$$\mathbf{r}_i = \mathbf{a}_i + \sum_\mu z_{i\mu} \mathbf{J}_\mu \tag{17}$$

where each component of $\mathbf{J}_\mu \in \mathcal{R}^3$ is an independent Gaussian noise variable of unit variance.

Similarly, we have for a subchain

$$\mathbf{r}_\sigma = \sum_{i \in \sigma} \mathbf{r}_i \equiv \mathbf{a}_\sigma + \sum_\mu z_{\sigma\mu} \mathbf{J}_\mu, \tag{18}$$

where $\mathbf{a}_\sigma = \sum_{i \in \sigma} \mathbf{a}_i$ and $\mathbf{z}_\sigma = \sum_{i \in \sigma} \mathbf{z}_i$. Thus, the noise amplitudes are additive.

The matrix inverse of $G$ can be similarly decomposed:

$$G^{-1}_{ij} = \mathbf{w}_i \cdot \mathbf{w}_j \tag{19}$$

where $w_{i\mu}$ is the (transposed) matrix inverse of $z_{i\mu}$:

$$\mathbf{z}_i \cdot \mathbf{w}_j = \delta_{ij} \tag{20}$$

Note that $\mathbf{z}_i$, $\mathbf{w}_i$ and $\mathbf{z}_\sigma$ are vectors in $R^{N-1}$, not in $R^3$.

In terms of $\mathbf{a}_i$ and $\mathbf{z}_i$, the variational free energy ignoring trivial additive constants (see ref. [30] for details) is given by

$$\begin{aligned}\hat{F} &= -3T \ln \det z + \frac{1}{2} \sum_i (3\mathbf{z}_i^2 + \mathbf{a}_i^2) \\ &+ \sum_\sigma \frac{1}{2a_\sigma} \exp\left(-\frac{a_\sigma^2}{2z_\sigma^2}\right) \left\{ \Psi\left(\kappa z_\sigma - \frac{a_\sigma}{z_\sigma}\right) - \Psi\left(\kappa z_\sigma + \frac{a_\sigma}{z_\sigma}\right) \right\}\end{aligned} \tag{21}$$

where

$$\Psi(x) \equiv \exp(x^2/2) \, \mathrm{erfc}(x/\sqrt{2}) \tag{22}$$

Setting $\mathbf{a}_i = 0$ for the unscreened case ($\kappa = 0$) the variational free energy simplifies to

$$\hat{F} = -3T \ln \det z + \frac{3}{2} \sum_i \mathbf{z}_i^2 + \sqrt{\frac{2}{\pi}} \sum_\sigma \frac{1}{z_\sigma} \tag{23}$$

which very much resembles the energy of an $(N-1)$-dimensional Coulomb chain with bonds $\mathbf{z}_i$, but with an extra *entropy* term (the first) preventing alignment of the ground state.

The equations for a local extremum of $\hat{F}(\mathbf{a}, \mathbf{z})$ are obtained by differentiation with respect to $\mathbf{z}_i$ and $\mathbf{a}_i$,

$$\frac{\partial \hat{F}}{\partial \mathbf{z}_i} = 0 \,, \quad \frac{\partial \hat{F}}{\partial \mathbf{a}_i} = 0 \tag{24}$$

Due to the use of relative coordinates and of local noise amplitudes, a simple gradient descent method with a large step-size $\epsilon$ can be used, that gives fast convergence to a solution of eqs. (24)

$$\Delta \mathbf{z}_i = -\epsilon_z \frac{\partial \hat{F}}{\partial \mathbf{z}_i}, \ \Delta \mathbf{a}_i = -\epsilon_a \frac{\partial \hat{F}}{\partial \mathbf{a}_i}, \tag{25}$$

Further speed is gained by updating the reciprocal variables $\mathbf{w}_i$ (arising from differentiating $\ln \det z$) using incremental matrix inversion (the Sherman-Morrison method [34, 30]) – the increment in $\mathbf{w}_j$ due to a change $\Delta \mathbf{z}_i$ is given by

$$\Delta \mathbf{w}_j = -\frac{\mathbf{w}_i(\mathbf{w}_j \cdot \Delta \mathbf{z}_i)}{1 + \mathbf{w}_i \cdot \Delta \mathbf{z}_i} \tag{26}$$

The resulting computational demand of the method is $\propto N^3$.

The high and low T properties of the variational approach were analyzed in ref. [30]. In contrast to the case in MC simulations, the free energy is directly accessible with the variational method. Furthermore, the approach respects the virial identity, $2\langle E_G \rangle - \langle E_C \rangle = 3(N-1)T$, in the sense that it holds also for the corresponding variational averages.

## 3.2 Non-Titrating Variational Results Revisited

The method, restricted to $\mathbf{a}_i = 0$, was extensively confronted with MC results in ref. [30]. Very good agreement was found for configurational quantities in the case of an unscreened Coulomb interaction (the error is well within the theoretical 11% limit [30]). In the screened case the method does not reproduce the MC results equally well although it gives a qualitatively correct picture of conformational properties.

Prior to dealing with the titration case we will augment the comparisons in [30] on the end-to-end distance $r_{ee}$ with those arising from the $\mathbf{a}_i \neq 0$ solutions and also with those produced by the simplest possible variational Ansatz [35]. The latter is a highly constrained version of eq. (14), with $\mathbf{z}_i = 0$ and constant $\mathbf{a}_i = \mathbf{a}$, corresponding to a *rigid rod*. This simplification leads to simple scaling behaviour [35],

$$r_{ee} \approx N(\ln N)^{1/3} \tag{27}$$

which seems to be approximately correct in MC calculations [30]; it is certainly correct at $T \to 0$. In fig. 1 we compare the results of the three different variational approaches ($\mathbf{a}_i = 0$, $\mathbf{a}_i \neq 0$, and the rigid rod) to MC data. The large $N$ behaviour of the $\mathbf{a}_i = 0$ and $\mathbf{a}_i \neq 0$ variational curves is consistent with approaching the theoretical limits derived in ref. [29]; 11% and 0% respectively. The average monomer-monomer distances are also very well reproduced by both the $\mathbf{a}_i = 0$ and $\mathbf{a}_i \neq 0$ variational approaches.

The simple rigid rod approach is more or less tailored to reproduce $r_{ee}$, but it lacks the degrees of freedom to properly describe local properties along the chain. A similar approach was pursued in ref. [8] in which good results were reported, at temperatures significantly higher than those at which the calculations in this paper are performed. We expect the accuracy of the rigid rod approximation to increase with decreasing temperature.

For screened Coulomb chains the $\mathbf{a}_i \neq 0$ results do not compare favourably with MC data – the chain tends to elongate more than the screened potentials call for. The same is true for the rigid rod approximation.

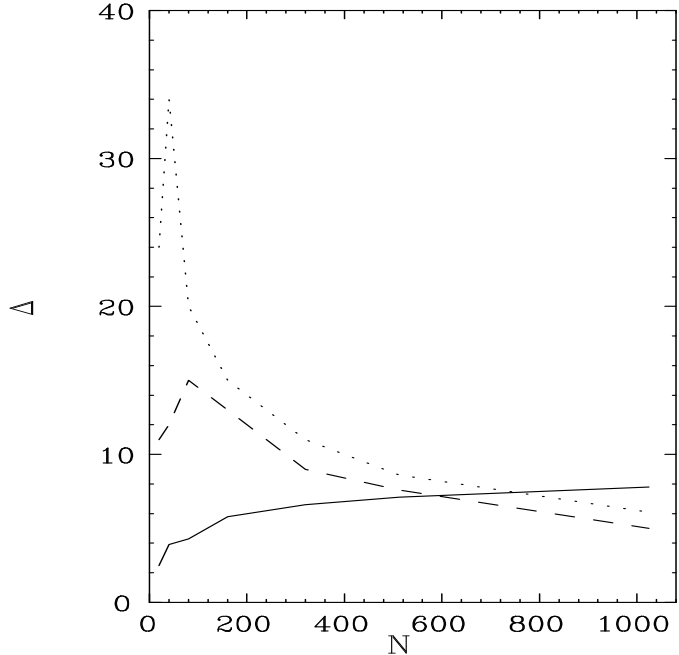

Figure 1: Relative difference ($\Delta$=(Var-MC)/MC) in % for $r_{ee}$ between different variational versions and MC data for unscreened Coulomb chains as functions of $N$. The $\mathbf{a}_i = 0$, $\mathbf{a}_i \neq 0$, and the rigid rod variants of the variational method are denoted by full, dashed and dotted lines respectively.

## 4 Variational Approach to Titration

### 4.1 Variational Treatment of Titratable Charges

So far the variational approach has been confined to the situation with fixed identical charges along the chain. Next we generalize to allowing charge exchange with the solvent, with the total charge governed by a chemical potential. This amounts to considering $s_i$ in eq. (8) as dynamical variables, that each can be either 0 or 1. Thus, suppressing for a moment the coordinate degrees of freedom, the system is isomorphic to an Ising spin system.

$$E = -\frac{1}{2}\sum_{ij} w_{ij} s_i s_j + \mu \sum_i s_i \qquad (28)$$

where the chemical potential $\mu$ has been introduced and coordinate dependencies etc. are lumped into the "couplings" $w_{ij}$. Such systems have been subject to much attention in the solid state community, in particular for describing magnetic properties of so called spin glasses. A powerful alternative to computing thermodynamical properties of eq. (28) by means of MC techniques is the mean field (MF) approximation. This can be considered as a variational approach along the lines above, with the variational energy Ansatz

$$E_V = -T \sum_i u_i s_i \tag{29}$$

where the coefficients $u_i$ are the variational parameters. Minimizing the corresponding variational free energy one gets the MF equations

$$v_i = g(u_i) \equiv \frac{1}{2}\left(1 + \tanh\left(\frac{u_i}{2}\right)\right) \tag{30}$$

$$u_i = -\frac{\partial E}{\partial v_i}\frac{1}{T} \tag{31}$$

which can be solved by iteration. The *mean fields* $v_i$ have the interpretation of mean charges $\langle s_i \rangle_V$, while $u_i$ are conventionally referred to as *local fields*.

We will next merge the variational treatments of relative coordinates and charges treating the unscreened and screened cases separately. For simplicity we limit the presentation to the $\mathbf{a}_i = 0$ solutions. The corresponding expressions for the $\mathbf{a}_i \neq 0$ solutions can be found in Sect.2.

## 4.2 Unscreened Coulomb Chain

For an unscreened Coulomb chain, the energy expression of eq. (8) simplifies to

$$E = \frac{1}{2}\sum_i r_i^2 + \sum_\sigma \frac{s_{l_\sigma} s_{r_\sigma}}{r_\sigma} + \mu \sum_i s_i \tag{32}$$

where $s_{l_\sigma}$ and $s_{r_\sigma}$ are the charges at the left and right endpoints of the sub-chain $\sigma$, respectively. With a variational effective energy Ansatz

$$E_V/T = \frac{1}{2}\sum_{ij} G_{ij}^{-1} \mathbf{r}_i \cdot \mathbf{r}_j - \sum_i u_i s_i \tag{33}$$

including both coordinate and charge degrees of freedom, the variational free energy becomes, ignoring additive constants

$$\hat{F} = -3T \ln \det z + T \sum_i \{u_i v_i - \ln(1+e^{u_i})\} + \frac{3}{2}\sum_i \mathbf{z}_i^2 + \sqrt{\frac{2}{\pi}}\sum_\sigma \frac{v_{l_\sigma} v_{r_\sigma}}{z_\sigma} + \mu \sum_i v_i \tag{34}$$

where $v_i$ and $u_i$ are related according to eq. (30); the first two terms give the entropy part $-TS_V$, while the remaining terms correspond to $\langle E \rangle_V$. The equations for a local extremum of $\hat{F}(\mathbf{z}, \mathbf{v})$ are obtained by differentiation with respect to $\mathbf{z}_i$ and $v_i$. One gets

$$-3T\mathbf{w}_i + 3\mathbf{z}_i - \sqrt{\frac{2}{\pi}}\sum_{\sigma \ni i} \frac{v_{l_\sigma} v_{r_\sigma} \mathbf{z}_\sigma}{z_\sigma^3} = 0 \tag{35}$$

and

$$\mu + Tu_i + \sqrt{\frac{2}{\pi}}\sum_{l \neq i} \frac{v_l}{z_{il}} = 0 \tag{36}$$

respectively.

## 4.3 Screened Coulomb Chain

The variational free energy for the screened case modifies to

$$\hat{F} = -3T \ln \det z + T \sum_i \{u_i v_i - \ln(1 + e^{u_i})\} \tag{37}$$

$$+ \frac{3}{2} \sum_i \mathbf{z}_i^2 + \sum_\sigma v_{l_\sigma} v_{r_\sigma} \left( \sqrt{\frac{2}{\pi}} \frac{1}{z_\sigma} - \kappa \Psi(\kappa z_\sigma) \right) + \mu \sum_i v_i$$

The derivatives then become

$$\frac{\partial \hat{F}}{\partial \mathbf{z}_i} \equiv -3T\mathbf{w}_i + 3\mathbf{z}_i - \sum_{\sigma \ni i} \frac{v_{l_\sigma} v_{r_\sigma} \mathbf{z}_\sigma}{z_\sigma^3} \left\{ \sqrt{\frac{2}{\pi}}(1 - \kappa^2 z_\sigma^2) + \kappa^3 z_\sigma^3 \Psi(\kappa z_\sigma) \right\} = 0 \tag{38}$$

$$\frac{\partial \hat{F}}{\partial v_i} \equiv \mu + T u_i + \sum_{l \neq i} v_l \left( \sqrt{\frac{2}{\pi}} \frac{1}{z_{li}} - \kappa \Psi(\kappa z_{li}) \right) = 0 \tag{39}$$

## 4.4 Properties of the Variational Solution

**The Apparent Equilibrium Constant**

The variational approximation to $\alpha$ is simply given by

$$\alpha = \frac{1}{N} \sum v_i = \overline{v} = \overline{g(u)} \tag{40}$$

The local fields $u_i$ can be expressed in terms of $v_i$ (inverting eq. (30)) as

$$u_i = g^{-1}(v_i) = \ln \frac{v_i}{1 - v_i} \tag{41}$$

Thus, the variational $\Delta pK$ is given by

$$\Delta pK \ln 10 = -\frac{\mu}{T} - g^{-1}\left(\overline{g(u)}\right) \approx -\frac{\mu}{T} - \overline{u} \tag{42}$$

In the unscreened case (for simplicity), the last expression amounts to

$$\frac{1}{NT} \sqrt{\frac{2}{\pi}} \sum_i \sum_{j \neq i} \frac{v_j}{z_{ij}} = \frac{1}{NT} \sum_i \sum_{j \neq i} \left\langle \frac{s_j}{r_{ij}} \right\rangle_V \tag{43}$$

where the interpretation of $\Delta pK$ as an average energy cost per dissociated charge is very clear.

**The Virial Identity**

In the unscreened case, there is, also in the titrating version, a virial identity,

$$2\langle E_G \rangle - \langle E_C \rangle = 3(N-1)T \tag{44}$$

obeyed by the exact thermodynamic ensemble. By taking the scalar product of eq. (35) with $\mathbf{z}_i$ and summing over $i$, we obtain

$$-3T(N-1) + 3\sum_i \mathbf{z}_i^2 - \sqrt{\frac{2}{\pi}} \sum_\sigma \frac{v_{l_\sigma} v_{r_\sigma}}{z_\sigma} = 0 \qquad (45)$$

and by a comparison with eq. (34), it is seen that the virial identity is indeed satisfied also for the variational expectation values. The same goes for the $\mathbf{a}_i \neq 0$ solutions.

**Structure of the Effective Energy**

By taking the scalar product of eq. (35) with $\mathbf{w}_j$, we obtain the effective force-constants

$$TG_{ij}^{-1} = \delta_{ij} - \frac{1}{3}\sqrt{\frac{2}{\pi}} \sum_{\sigma \ni i,j} \frac{v_{l_\sigma} v_{r_\sigma}}{z_\sigma^3} \qquad (46)$$

while eq. (36) gives the effective local fields

$$-Tu_i = \mu + \sqrt{\frac{2}{\pi}} \sum_{l \neq i} \frac{v_l}{z_{il}} \qquad (47)$$

Thus, we can rewrite the variational energy, eq. (33), as

$$E_V(\{\mathbf{r}_i, s_i\}) = \frac{1}{2}\sum_i \mathbf{r}_i^2 + \sum_\sigma \sqrt{\frac{2}{\pi}} \left( \frac{v_{l_\sigma} s_{r_\sigma} + v_{r_\sigma} s_{l_\sigma}}{z_\sigma} - \frac{1}{6}\frac{v_{l_\sigma} v_{r_\sigma} \mathbf{r}_\sigma^2}{z_\sigma^3} \right) + \mu \sum_i s_i \qquad (48)$$

where the first and the last term reproduce the true bond and chemical potential terms of eq. (32), while the middle term emulates the effect of the Coulomb interaction on the charges (by appropriate local fields), and on the conformation (by suitable repulsive spring forces).

The variational optimization thus forces the effective energy to have an intuitively appealing (and indeed very reasonable) structure.

## 4.5 Algorithm Implementation

For the numerical minimization of the free energy of eqs. (34) or (37) with respect to the variational parameters $z_i$ ($\mathbf{w}_i$) and $v_i$ ($u_i$) a modified gradient descent is used;

$$\Delta z_i = -\epsilon_z \frac{\partial \hat{F}}{\partial \mathbf{z}_i}, \quad \Delta u_i = -\epsilon_u \frac{\partial \hat{F}}{\partial v_i}. \qquad (49)$$

where the use of the $v$ (rather than $u$) derivative in the $u$-update implies a dynamical step size

$$\tilde{\epsilon}_u = \frac{\epsilon_u}{v(1-v)} \qquad (50)$$

which is found to speed up the calculations. The $\mathbf{w}_i$:s are obtained using incremental matrix inversion (eq. (26)).

The complete algorithm looks as follows:

1. Initialize all $z_i$ and $v_i$ at random.

2. Repeat until convergence:

   For all i:

   - Update $z_i$ (if $i < N$) and $v_i = g(u_i)$ according to eq. (49).
   - Correct all $w_j$ according to eq. (26)

3. Extract $G_{ij} = z_i \cdot z_j$ and compute variational averages of interest.

Typical step-sizes are $\epsilon_z \approx 0.15$ and $\epsilon_u \approx 0.5$. The number of of computations for each iteration step is proportional to $N^3$. An $N = 80$ system converges within about 100 iterations, which is somewhat slower than in the non-titrating case.

## 4.6 A Simplified Variational Approach

For a highly charged polyelectrolyte in the absence of salt, the conformation is strongly elongated. This fact motivates a simplified variational Ansatz, based on a rigid rod conformation and a constant degree of dissociation $\alpha$ along the chain. Such a simplified picture allows for analytical estimates of quantities of interest like $\Delta pK$ and $r_{ee}$. To this end, consider the non-fluctuating limit $z_i \to 0$ with $a_i = a = R/(N-1)$ for all $i$, in which the variational Boltzmann distribution is given by

$$\exp(-E_V/T) = \exp\left(u \sum s_i\right) \prod_i \delta\left(r_i - a\right) \quad (51)$$

Identifying $\alpha$ with $g(u)$ gives for large $N$ the variational free energy

$$\hat{F} = NT\{\alpha \ln(\alpha) + (1-\alpha)\ln(1-\alpha)\} + \frac{R^2}{2(N-1)} + \frac{\alpha^2}{R}N(N-1)(\ln N + \gamma - 1) + N\mu\alpha \quad (52)$$

Here the sum $\sum_1^N 1/k$ has been approximated by $\ln N + \gamma$, where $\gamma$ is the Euler constant. Minimizing $\hat{F}$ with respect to $\alpha$ and $R$ gives for large N

$$r_{ee} = R \approx \alpha^{2/3} N (\ln N)^{1/3} \quad (53)$$

$$\langle E_C \rangle \approx \alpha^{4/3} N (\ln N)^{2/3} \quad (54)$$

$$\Delta pK \approx \frac{2}{T \ln 10} \alpha^{1/3} (\ln N)^{2/3} \quad (55)$$

where $\mu$ has been eliminated in favour of $\alpha$. This simplified model should give good results in the highly charged limit $\alpha \to 1$, whereas its lack of fluctuations should give rise to poor performance in the $\alpha \to 0$ limit.

A similar approximation can of course be used for a screened Coulomb potential. In this case, however, there is no analytical solution, but the relevant equations are readily solved by a simple numerical iteration scheme.

# 5 Monte Carlo Methods

The Monte Carlo (MC) simulations were performed with the traditional Metropolis algorithm [36] in a semi-grand canonical ensemble. A single polyelectrolyte chain was simulated with the charges/protons moving between the monomers and an implicit bulk of fixed chemical potential.

When a proton move is attempted, a monomer is picked at random and the charge state of the monomer is switched. The associated (free) energy change, $\Delta E$, which determines if the move is to be accepted or rejected according to the Metropolis scheme, is the sum of the change $\Delta E_C$ in the intramolecular Coulomb energy and a term $\pm\mu$ that corresponds to the change in free energy for the acid-base reaction of an isolated monomer; the minus sign is used when the monomer is to be protonated, and the plus sign when it is to be deprotonated.

In a single MC step a proton is only moved from the chain to the bulk or *vice versa*. Adding a step where a proton moves within the polymer does not affect the averages but increases the calculation time.

When a conformational change is attempted, the associated energy change is given by the change in bond energy, $\Delta E_G$, plus the electrostatic energy change, $\Delta E_C$. The sampling of chain conformations is made highly efficient by using a pivot algorithm, which allows chain lengths of more than 2000 monomers. The pivot algorithm was first described by Lal [37], and its efficiency for self-avoiding walks has been thoroughly discussed by Madras and Sokal [38].

A traditional move to update the coordinate variables is to attempt a translation of only one monomer at a time. The number of interactions that have to be calculated is of the order $N$ for a highly charged chain and a large number of attempts per monomer is needed to generate independent chain conformations. In the pivot algorithm, however, each monomer $i$ (except the first one) is translated in turn but together with the remaining semi-chain (monomers $i+1$ to $N$). Furthermore, the semi-chain is then rotated as a rigid body around one of the coordinate axes with monomer $i$ as origin. The number of interactions calculated in one step is of the order $N^2$ but independent conformations are obtained after only a few attempted moves, on the order of one per monomer or $N$ in total. The net effect is a greatly reduced simulation time for a given degree of precision and a computational cost that grows approximately as $N^3$. A completely different and even slightly more efficient procedure has recently been described by Irbäck [39].

A change in conformation is attempted once in every 20 steps; in the remaining steps, a change in the charge state is attempted. The total number of steps is around $10^8$. Every run is preceded by an equilibration of $10^5$–$10^6$ steps, where a change in conformation is attempted every other step, starting from a straight line with a charge on every other monomer. The simulations are faster at high $\tilde{\mu}$, since fewer monomers are charged and thus the number of interactions that have to be calculated is smaller.

# 6 Results and Discussion

## 6.1 Unscreened Coulomb Chain

We now compare the variational and Monte Carlo results for $r_{ee}$ and $\Delta pK$ as a function of chain length and degree of dissociation. Fig. 2 shows an excellent agreement between the $\mathbf{a}_i = 0$ variational results for $r_{ee}$ and MC data, with a maximal relative error of 8% for the $N = 1000$ chain at $\alpha = 1$, well within the theoretical high-$N$ limit of 11%. The difference between the variational and MC results increases with $\alpha$ as expected, but is indeed satisfactory for all cases studied. The $\mathbf{a}_i \neq 0$ variational solution is slightly inferior to the purely fluctuating solution, except for the $N = 1000$ curve at high degree of ionization. Thus, we find a qualitative behaviour of the two variational solutions similar to that seen in fig. 1.

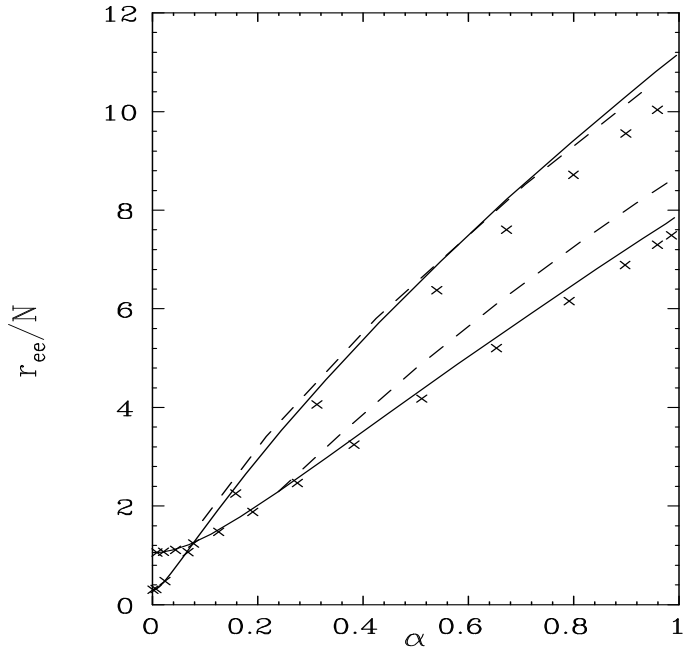

Figure 2: $\tilde{r}_{ee}/N$ as a function of $\alpha$ for for an unscreened polyelectrolyte chain. Monte Carlo results are denoted by crosses and the $\mathbf{a}_i = 0$ and $\mathbf{a}_i \neq 0$ variational solutions are drawn as solid and dashed lines, respectively. The upper curves are for $N = 1000$ and the lower for $N = 80$ and the unit of length is Å.

The rigid rod approximation suggests that $r_{ee}/N$ should increase linearly with $\alpha^{2/3}(\ln N)^{1/3}$, which seems to be verified in fig. 3. Similar scaling behaviour for an analogous non-titrating polyelectrolyte has been derived in the literature [35, 7, 8]. We have, however, made numerical estimates of the exponents from logarithmic plots of $r_{ee}/N$ against $\alpha$ from both variational and MC data. For the $\alpha$ dependence we find exponents in the range 0.80-0.85 from both approaches, which is significantly larger than the value of 2/3 predicted by the rigid rod approximation. The origin of the discrepancy is unclear. A similar discrepancy is seen for the $\ln N$ exponent, which in the rigid rod case is given by 1/3, while numerical estimates from variational and MC data suggest a value of 0.6-0.7. One possible explanation could be that the chain expands via two different mechanisms. One is the expansion of each monomer bond, which should give rise to an $\ln N$ exponent of 1/3 [30]; Another is the increasing alignment of monomer bonds with $N$, which reaches its limit when the coupling becomes strong and should level off for large $N$.

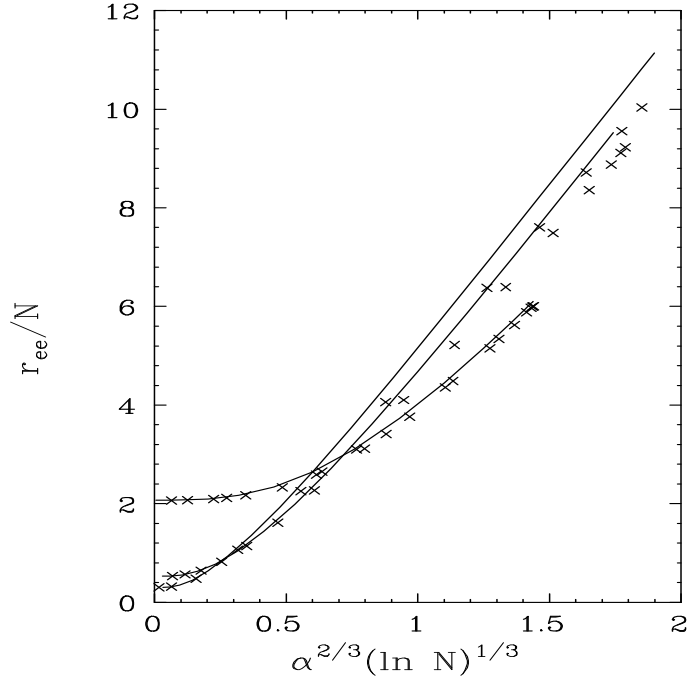

Figure 3: $\tilde{r}_{ee}/N$ as a function of $\alpha^{2/3}(\ln N)^{1/3}$ for different chain lengths ($N = 20, 320, 1000$) in the case of an unscreened Coulomb potential. Variational ($\mathbf{a}_i = 0$) and MC results are denoted by solid lines and crosses, respectively. The unit of length is Å.

Fig. 4 shows the shift in the apparent dissociation constant upon ionization of the chain and one finds that the $\mathbf{a}_i = 0$ variational results differ significantly from the MC results for the longer chains. The $\mathbf{a}_i \neq 0$ results, on the other hand, are always in excellent agreement with the MC data with negligible differences for all systems studied. The largest error seen is of the order of one tenth of a $pK$ unit.

The global conformational properties of the polymer depends on the electrostatic coupling strength via an *effective coupling*, approximately given by

$$\frac{\alpha^{4/3} N}{T} \tag{56}$$

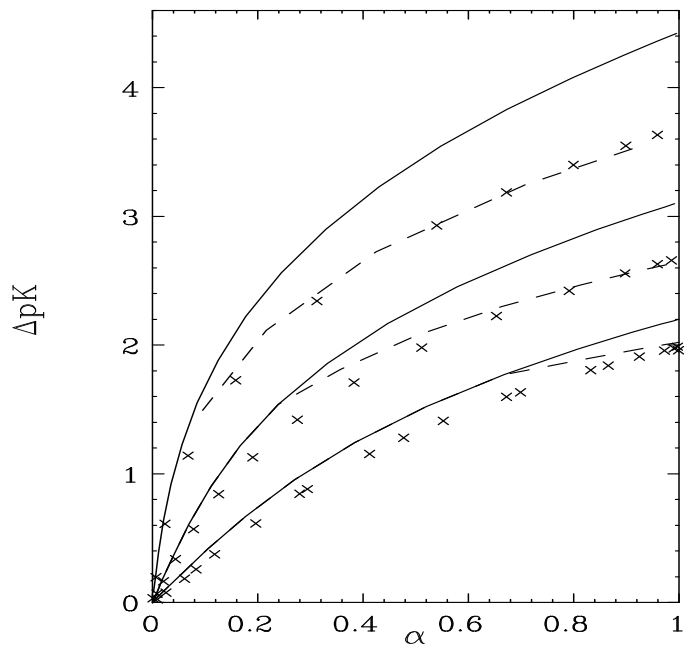

Figure 4: $\Delta pK$ as a function of $\alpha$ for different chain lengths with an unscreened Coulomb potential. Monte Carlo results are denoted by crosses and the $\mathbf{a}_i = 0$ and $\mathbf{a}_i \neq 0$ variational solutions are drawn as solid and dashed lines, respectively. The chain lengths are from top to bottom, $N = 1000, 80, 20$.

The two variational solutions coincide for a weakly coupled chain, while at higher coupling two distinct solutions appear. This is seen in fig. 4 where the $\mathbf{a}_i \neq 0$ solution appears at a lower $\alpha$ value for the longer chains. Lowering the temperature (or decreasing $r_0$) would have a similar effect.

The asymptotic behaviour in the rigid rod approximation can be investigated by plotting $\Delta pK$ against $\alpha^{1/3}(\ln N)^{2/3}$. Fig. 5 shows that the $\alpha$ dependence is not at all well described by the rigid rod over the parameter range studied and that it is only for very large $N$ and $\alpha$ close to unity, that the slope seems to approach $1/3$. A similar plot can be made for the $\ln N$ dependence with a slightly better agreement and with a linear relation between $\ln \Delta pK$ and $\ln(\ln N)$. The slope is approximately 0.7 for $\alpha = 1.0$, in good agreement with the rigid rod prediction of $2/3$, but it increases with decreasing $\alpha$. This is again consistent with two different expansion mechanisms.

The rigid rod approximation gives an excellent description of the titration behaviour for a highly charged chain, while it slightly deviates from the MC results for short chains at low $\alpha$. Fig. 6 shows that the rigid rod actually gives a better approximation to the MC data for $\Delta pK$ at high $\alpha$ values than the full $\mathbf{a}_i = 0$ variational solution. A similarly good agreement is found in fig. 7 where the MC and rigid rod numbers are virtually indistinguishable. The agreement between the rigid rod and the MC results holds for $r_{ee}$ and the apparent dissociation constant, but not for local properties like the monomer-monomer separation. The rigid rod does not distinguish between different positions along the chain and it also strongly underestimates the average monomer-monomer separation. For such properties, the more sophisticated variational approaches are definitely superior and in general in good agreement with the exact MC results.

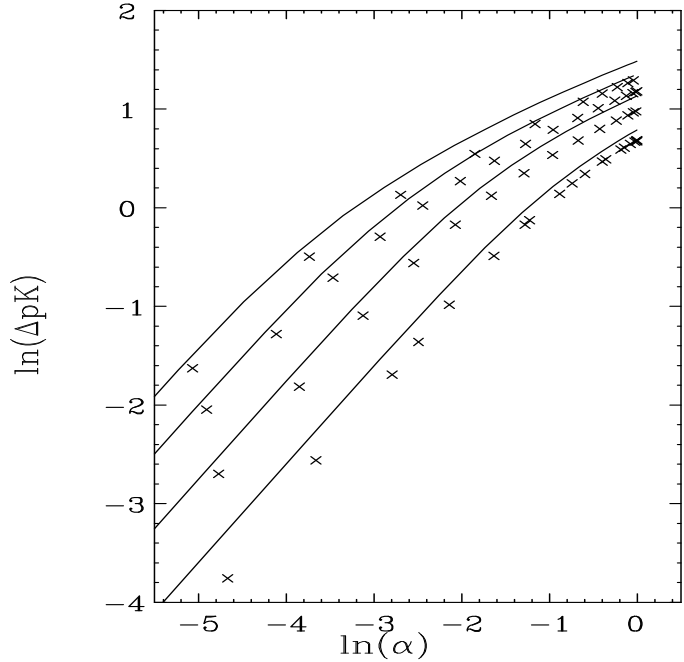

Figure 5: $\ln \Delta pK$ as a function of $\ln \alpha$ for unscreened polyelectrolyte chains. From bottom to top: $N = 20, 80, 320$, and $1000$. Variational ($\mathbf{a}_i = 0$) and MC results are denoted by solid lines and crosses, respectively.

## 6.2 Screened Coulomb Chain

A discussion of the accuracy of the screened Coulomb approximation is beyond the scope of the present communication [4, 25]. Our principal aim is to investigate the variational techniques and we will use the screened Coulomb potential assuming that it contains the main physical features relating to screening. We will be primarily interested in $r_{ee}$ and $\Delta pK$. These quantities will be functions of chain length and degree of ionization as before, but they will also depend on the screening parameter $\kappa$. In a real solution $\kappa$ will contain contributions from added salt as well as any other charged molecule like the polyelectrolyte chain itself. Here we will neglect all these complications and we will refer to different screening conditions by stating the corresponding univalent salt concentration.

A polyelectrolyte chain in a salt solution will be approximately independent of electrolyte concentration as long as the screening length is larger than the end-to-end separation and the screening will start to play a role when $\kappa^{-1}$ is smaller than or of the same order as $r_{ee}$.

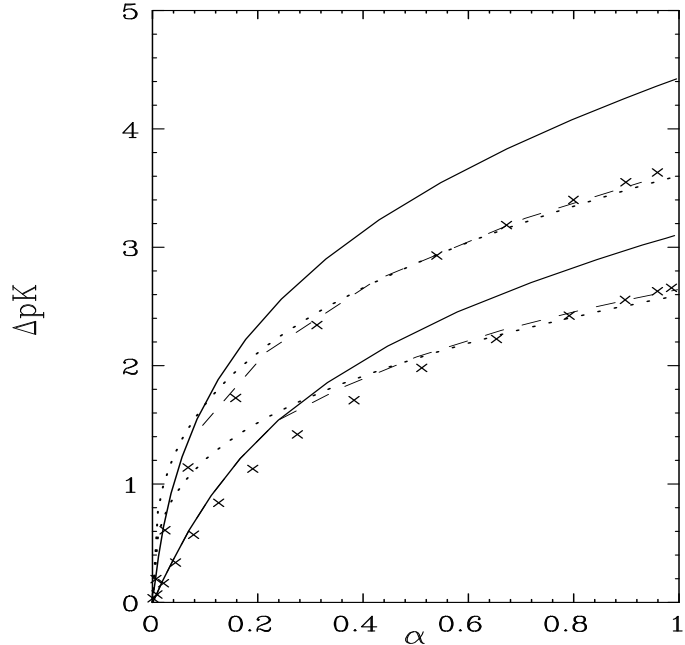

Figure 6: $\Delta pK$ as a function of $\alpha$ for an unscreened Coulomb chain. Monte Carlo results are denoted by crosses and the $\mathbf{a}_i = 0$, $\mathbf{a}_i \neq 0$ and the rigid rod solutions are drawn as solid, dashed and dotted lines, respectively. The upper curves are for $N = 1000$ and the lower for $N = 80$.

The accuracy of the variational result for $r_{ee}$ from the purely fluctuating solution ($\mathbf{a}_i = 0$) deteriorates with increasing chain length for a given salt concentration. The discrepancy to the MC data also becomes worse with increasing degree of ionization - see fig. 8. With increasing salt concentration, the accuracy will deteriorate up to some value, whereafter it improves. In the limit of very high screening the chain becomes perfectly Brownian, leading to an exact agreement. The variational predictions for $r_{ee}$ consistently overshoots; this can be attributed to an overestimate of the interaction with a Gaussian Boltzmann distribution.

The variational solution with $\mathbf{a}_i \neq 0$ is absent at low $\alpha$ and high salt concentration. In the other limit there will always be two solutions and we find that the purely fluctuating one is the most accurate one in predicting the end-to-end separation. The $\mathbf{a}_i \neq 0$ solution on the other hand predicts much too large $r_{ee}$ - see fig. 8. The failure to describe the global structure of a screened chain is mainly due to the alignment of the bond vectors $\mathbf{a}_i$, which leads to the wrong asymptotic behaviour. The chain will expand like $N$ instead of $N^{3/5}$, which approximately seems to describe the MC results. The rigid rod approximation for the screened chain will of course show a behaviour for the end-to-end separation similar to that of the full $\mathbf{a}_i \neq 0$ solution.

The apparent dissociation constant in the presence of salt (see fig. 9) is not well described by the fluctuating solution. Instead we find the $\mathbf{a}_i \neq 0$ solution, despite its obvious structural failure, to be superior and in fair agreement with MC data. Obviously the coupling between thermodynamic derivatives, like the apparent dissociation constant, and the global structure is rather weak, and it is the local structure and fluctuations that determine the thermodynamics of a screened chain. The relative error is of the same size as for the unscreened chain - typically of the order of 10% or less. We find that $\Delta pK$ is consistently too large in the variational approximations, in accordance with the variational overestimate of the interaction.

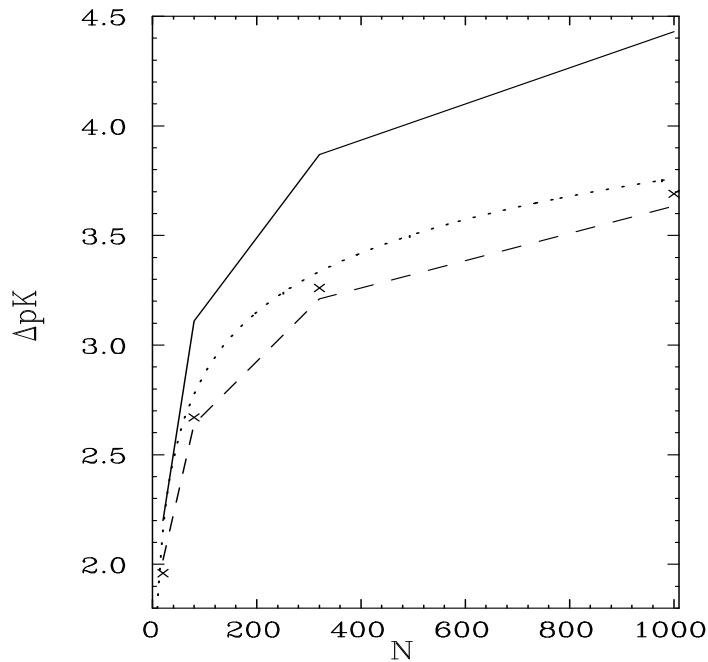

Figure 7: $\Delta pK$ as function of $N$ for $\alpha = 1.0$ for an unscreened polyelectrolyte chain. Monte Carlo results are denoted by crosses and the $\mathbf{a}_i = 0$, $\mathbf{a}_i \neq 0$ and the rigid rod solutions are drawn as solid, dashed and dotted lines, respectively.

For the rigid rod approximation, the overall agreement with MC data is not as good as for the unscreened chain, and $\Delta pK$ is predicted to be essentially independent of $\alpha$. However, as seen in fig. 9, the numerical values obtained for $\alpha = 1$ agree very well with simulated numbers, and the salt dependence of $\Delta pK$ for the fully ionized chain is accurately reproduced.

# 7  Conclusions

The variational calculations accurately reproduce structural and thermodynamic properties of a titrating polyelectrolyte interacting via an unscreened Coulomb potential. For a highly charged polyelectrolyte we find two distinct solutions to the Gaussian variational Ansatz, which at weak coupling collapse into one – a purely fluctuating solution with $\mathbf{a}_i = 0$. At strong effective coupling – i.e. for large $N$, low $T$ and large $\alpha$ – the $\mathbf{a}_i \neq 0$ solution always gives the lowest free energy, as well as the best approximation to structural and thermodynamic properties. At intermediate coupling strength, the purely fluctuating solution can produce superior structural data, while the apparent dissociation constant seems to be best described by the $\mathbf{a}_i \neq 0$ solution.

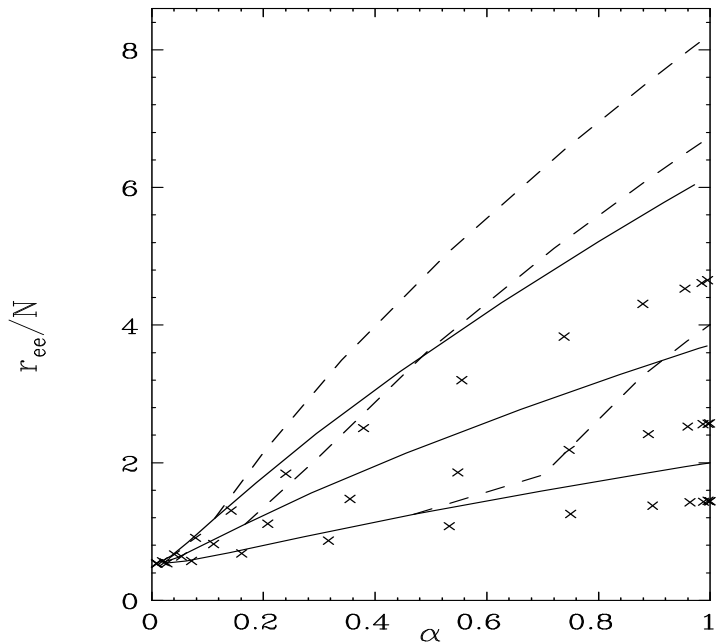

Figure 8: $\tilde{r}_{ee}/N$ as a function of $\alpha$ for a screened Coulomb chain with $N = 320$. Monte Carlo results are denoted by crosses and the $\mathbf{a}_i = 0$ and $\mathbf{a}_i \neq 0$ solutions are drawn as solid and dashed lines, respectively. The salt concentrations are from top to bottom 0.001, 0.01 and 0.1 M and the unit of length is Å.

With increasing electrostatic coupling, the chain becomes stiffer and more rodlike and a less extensive variational approach, like the rigid rod, becomes applicable. This can be solved analytically, and turns out to be fairly accurate both for thermodynamics and for certain structural properties.

The screened Coulomb potential is more difficult to emulate with a Gaussian Ansatz, although reasonable predictions for the end-to-end separations are obtained from the purely fluctuating solutions. When the bond vectors $\mathbf{a}_i$ are non-zero they tend to align and the end-to-end separation increases linearly with $N$. For the purely fluctuating solution on the other hand, the end-to-end separation increases approximately as $N^{0.6}$ at high salt concentration, which is expected for a polymer with short range interactions. The apparent dissociation constant is well described also for a screened chain and we find that the $\mathbf{a}_i \neq 0$ solution, despite its structural shortcomings, is the best approximation for this property.

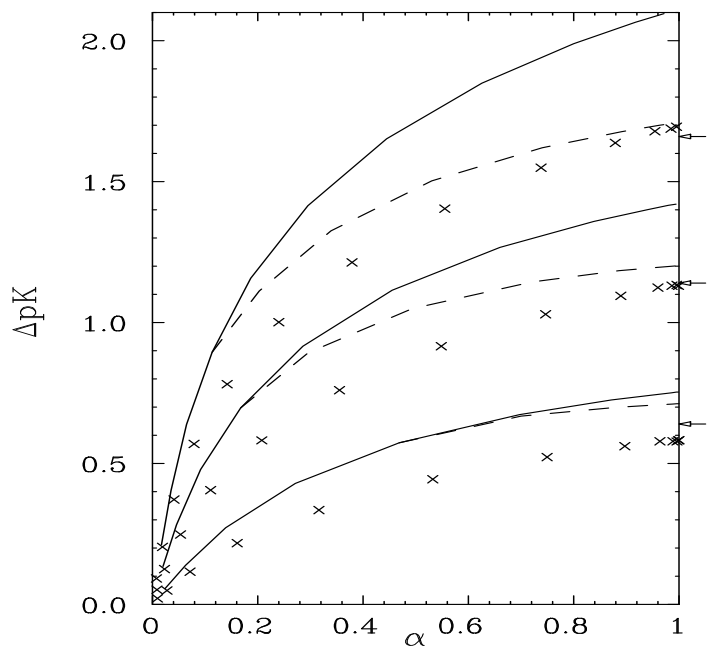

Figure 9: $\Delta pK$ as a function of $\alpha$ for a screened Coulomb chain with $N = 320$. Monte Carlo results are denoted by crosses and the $\mathbf{a}_i = 0$ and $\mathbf{a}_i \neq 0$ solutions are drawn as solid and dashed lines, respectively. The rigid rod results for $\alpha = 1$ are indicated with arrows. The salt concentrations are from top to bottom 0.001, 0.01 and 0.1 M.